\documentclass[a4paper,11pt]{article}

\usepackage[utf8]{inputenc}
\usepackage[affil-sl,auth-sc]{authblk}
\usepackage{amssymb,amsmath,amsbsy}
\usepackage{mathbbol}
\usepackage{bm}
\usepackage{appendix}
\usepackage[linktocpage=true]{hyperref}
\usepackage[top=1.3in, bottom=1.3in,left=0.9in,right=0.9in]{geometry}
\usepackage{graphicx}
\usepackage{cite}
\usepackage{float}
\usepackage{caption}
\usepackage{wasysym}
\usepackage{float}
\usepackage{slashed}
\usepackage{caption}
\usepackage{indentfirst}
\usepackage{multirow}
\usepackage{feynmp-auto}
\usepackage{braket}
\usepackage[nottoc]{tocbibind} % recommended to create references toc entry
\usepackage{color}

\newcommand{\vp}{\varphi}
\newcommand{\tvp}{\widetilde{\varphi}}

\newcommand{\vpj }{\mbox{${\vp^\dag
      i\,\raisebox{2mm}{\boldmath ${}^\leftrightarrow$}\hspace{-4mm}
      D_\mu\,\vp}$}} \newcommand{\vpjt}{\mbox{${\vp^\dag
      i\,\raisebox{2mm}{\boldmath ${}^\leftrightarrow$}\hspace{-4mm}
      D_\mu^{\,I}\,\vp}$}}
\let\originalleft\left \let\originalright\right
\renewcommand{\left}{\mathopen{}\mathclose\bgroup\originalleft}
\renewcommand{\right}{\aftergroup\egroup\originalright}
\numberwithin{equation}{section}
\newcommand\numberthis{\addtocounter{equation}{1}\tag{\theequation}}
\newcommand{\marrow}[5]{%
\fmfcmd{style_def marrow#1
expr p = drawarrow subpath (1/4, 3/4) of p shifted 6 #2 withpen pencircle scaled 0.4;
label.#3(btex #4 etex, point 0.5 of p shifted 6 #2); enddef;}
\fmf{marrow#1,tension=0}{#5}}
\newcommand{\fmfcounterterm}[1]{\fmfv{decor.shape=circle,decor.filled=0,decor.size=4thick,label=$\boldsymbol{\times}$,label.dist=0}{#1}}
\newcommand{\fmfblobbig}[1]{\fmfv{decor.shape=circle,decor.filled=50,decor.size=8thick}{#1}}
\newcommand{\fmfblobsmall}[1]{\fmfv{decor.shape=circle,decor.filled=50,decor.size=5thick}{#1}}
\newcommand{\fmfsquare}[1]{\fmfv{decor.shape=square,decor.filled=full,decor.size=3thick}{#1}}

\begin{document}
\begin{fmffile}{h2zg}

\title{\bf The decay $h\to Z \gamma$ in the Standard-Model
  \\ Effective Field Theory}

\author{Athanasios Dedes\footnote{email: {\tt adedes@uoi.gr}},~
  Kristaq Suxho\footnote{email: {\tt csoutzio@uoi.gr}} ~and~
  Lampros Trifyllis\footnote{email: {\tt ltrifyl@cc.uoi.gr}}}
\affil{\small Department of Physics, Division of Theoretical
  Physics, \\ University of Ioannina, GR 45110, Greece}

\date{\today}

\maketitle
\thispagestyle{empty}

\begin{abstract}
We calculate the $S$-matrix element for the Higgs boson decay to a $Z$-boson and
a photon, $h\to Z\gamma$, at one-loop in the Standard-Model Effective Field
Theory (SMEFT) framework and in linear $R_{\xi}$-gauges. Our SMEFT expansion
includes all relevant operators up to dimension-6 considered in Warsaw basis
without resorting to any flavour or CP-conservation assumptions. Within this
approximation there are 23 dimension-6 operators affecting the amplitude, not
including flavour and hermitian conjugation. The result for the on-shell $h\to
Z\gamma$ amplitude is gauge invariant, renormalisation-scale invariant and
gauge-fixing parameter independent. The calculated  ratio of the SMEFT versus
the SM expectation for the $h\to Z\gamma$ decay width is then written in a
semi-numerical form  which is useful for further comparisons with related
processes. 
For example, the $h\to Z\gamma$ amplitude contains 16 operators in common with
the $h\to \gamma\gamma$ amplitude and one can draw useful results about its
feasibility at current and future LHC data. 
\end{abstract}

\tableofcontents
%\listoffigures
%\listoftables
\newpage

%%%%%%%%%%%%%%%%%%%%%%%%%%%%%%%%%
\section{Introduction}
\label{sec:introduction}
%%%%%%%%%%%%%%%%%%%%%%%%%%%%%%%%%

The Higgs boson decay processes $h \to \gamma \gamma$ and $h \to Z \gamma$ are
extremely important probes for physics beyond the Standard Model (SM) and are
under intensive research ever since the Higgs boson discovery at
LHC~\cite{Aad:2012tfa,Chatrchyan:2012xdj}. 
Experimental bounds for both $h\to \gamma\gamma$ and $h\to Z\gamma$ decays were
set by the CMS and ATLAS collaborations at LHC
\cite{Chatrchyan:2013vaa,Aad:2014fia,Sirunyan:2018tbk,Aaboud:2017uhw}. Although
the $h\to \gamma\gamma$ decay width has been observed to within 15\% w.r.t.\ the
SM prediction, the situation is not the same for $h\to Z\gamma$. An upper bound
for $h\to Z\gamma$ given by ATLAS~\cite{Aaboud:2017uhw}, with center-of-mass
energy $\sqrt{s}=13\;\mathrm{TeV}$ proton-proton collisions, integrated
luminosity $36.1\;\mathrm{fb}^{-1}$, and Higgs boson mass
$M_{h}=125.09\;\mathrm{GeV}$, finds that $\sigma(pp \to h)\times B(h \to Z
\gamma)$ is $6.6$ times the SM prediction with 95\% confidence level. More
specifically, it is
\begin{equation}
\mu_{h\to Z\gamma} = \frac{\sigma(pp \to h)\times \mathrm{Br}(h \to Z \gamma)}
{\sigma(pp \to h)_{\mathrm{SM}}\times 
\mathrm{Br}(h \to Z \gamma)_{\mathrm{SM}}} \lesssim 6.6\,.
\label{mu}
\end{equation}
If physics beyond the SM does not affect the Higgs production,\footnote{We shall
comment upon this issue at the end of section \ref{sec:results}.} which mainly
goes via the gluon fusion process, $gg\to h$, then the bound of \eqref{mu} is
translated to a bound on a ratio
\begin{equation}
    \label{eq:RhZg}
    \mathcal{R}_{h\to Z\gamma} = 
    \frac{\Gamma(\mathrm{EXP},h\to Z\gamma)}{\Gamma(\mathrm{SM},h\to Z\gamma)}\,.
\end{equation}
The decay $h \to Z \gamma$ has been calculated for the first time in the SM in
refs.~\cite{Cahn:1978nz,Bergstrom:1985hp,Gunion:1987ke}.
To our knowledge, in the Standard-Model Effective Field Theory (SMEFT) this
process has been studied using a partial list of $d=6$ operators in
refs.~\cite{Ghezzi:2015vva,Cirigliano:2016nyn,Vryonidou:2018eyv}, while an
analysis with a complete set of $d=6$ operators has recently been performed  in
ref.~\cite{Dawson:2018pyl}. 
Here, we advance the current status of the SMEFT one-loop calculation
for $\mathcal{R}_{h\to Z\gamma}$ in eq.~\eqref{eq:RhZg} by presenting
\begin{itemize}
\item a clear and concise renormalisation framework in general $R_\xi$-gauges,
\item a gauge invariant master formula for the amplitude 
which self-explains several issues even 
for the SM-amplitude,
\item a semi-analytic formula for $\mathcal{R}_{h\to Z\gamma}$,
\item correlations between  the ratios $\mathcal{R}_{h\to Z\gamma}$ and 
$\mathcal{R}_{h\to \gamma\gamma}$.
\end{itemize}
Obviously there are many similarities in the calculation with the $h\to
\gamma\gamma$ decay worked out at one loop in SMEFT in
ref.~\cite{Dedes:2018seb}\footnote{For similar studies see also
refs.~\cite{Hartmann:2015aia,Hartmann:2015oia,Dawson:2018liq}.} and we follow
faithfully the renormalisation framework and the results found in there. We
shall only focus on technical aspects that arise strictly in calculating the
$h\to Z\gamma$ amplitude. This involves some subtle issues regarding gauge
invariance which we address in section~\ref{sec:renormalisation}. The operators
relevant for $h\to Z\gamma$ are discussed  in section~\ref{sec:operators} and
their effects in $\mathcal{R}_{h\to Z\gamma}$  in section~\ref{sec:results}. We
conclude in section \ref{sec:epilogue}.

%%%%%%%%%%%%%%%%%%%%%%
\section{Operators}
\label{sec:operators}
%%%%%%%%%%%%%%%%%%%%%

%%%%%%%%%%%%%%%%%%%%%%%%%%
\begin{table}[th] 
\centering
\renewcommand{\arraystretch}{1.3}
\begin{tabular}{||c|c||c|c||c|c||} 
\hline \hline
\multicolumn{2}{||c||}{$X^3$} & 
\multicolumn{2}{|c||}{$\vp^6$~ and~ $\vp^4 D^2$} &
\multicolumn{2}{|c||}{$\psi^2\vp^3$}\\
\hline
$ $                &  &
$Q_{\vp\Box}$ & $(\vp^\dag \vp)\raisebox{-.5mm}{$\Box$}(\vp^\dag \vp)$ &
$Q_{e\vp}$           & $(\vp^\dag \vp)(\bar l'_p e'_r \vp)$\\
$Q_W$          &  $\varepsilon^{IJK} W_\mu^{I\nu} W_\nu^{J\rho} W_\rho^{K\mu}$ &   
$Q_{\vp D}$   & $\left(\vp^\dag D^\mu\vp\right)^* \left(\vp^\dag D_\mu\vp\right)$ &
$Q_{u\vp}$           & $(\vp^\dag \vp)(\bar q'_p u'_r \tvp)$\\
$ $                &  &    
   &   &
$Q_{d\vp}$           & $(\vp^\dag \vp)(\bar q'_p d'_r \vp)$\\
\hline \hline
\multicolumn{2}{||c||}{$X^2\vp^2$} &
\multicolumn{2}{|c||}{$\psi^2 X\vp$} &
\multicolumn{2}{|c||}{$\psi^2\vp^2 D$}\\ 
\hline
$ Q_{\vp  W}$     &  $\vp^\dag \vp\,  W^I_{\mu\nu} W^{I\mu\nu}$  & 
$Q_{eW}$               & $(\bar l'_p \sigma^{\mu\nu} e'_r) \tau^I \vp W_{\mu\nu}^I$ &
${Q_{\vp l}^{(1)}}$      & ${(\vpj)(\bar l'_p \gamma^\mu l'_r)}$\\
$ Q_{\vp B} $         & $ \vp^\dag \vp\, B_{\mu\nu} B^{\mu\nu}$ &  
$Q_{eB}$        & $(\bar l'_p \sigma^{\mu\nu} e'_r) \vp B_{\mu\nu}$ &
${Q_{\vp l}^{(3)}}$      & ${(\vpjt)(\bar l'_p \tau^I \gamma^\mu l'_r)}$\\
$ Q_{\vp WB} $     & $ \vp^\dag \tau^I \vp\, W^I_{\mu\nu} B^{\mu\nu}$  & 
$ Q_{uW} $        &  $(\bar q'_p \sigma^{\mu\nu} u'_r) \tau^I \tvp\, W_{\mu\nu}^I$ &
${Q_{\vp e}}$            & ${(\vpj)(\bar e'_p \gamma^\mu e'_r)}$\\
$ $         &  &
$ Q_{uB} $               &  $(\bar q'_p \sigma^{\mu\nu} u'_r) \tvp\, B_{\mu\nu}$ &
${Q_{\vp q}^{(1)}}$      & ${(\vpj)(\bar q'_p \gamma^\mu q'_r)}$\\
$ $     &  &
$Q_{dW} $        & $(\bar q'_p \sigma^{\mu\nu} d'_r) \tau^I \vp\, W_{\mu\nu}^I$ &
${Q_{\vp q}^{(3)}}$      & ${(\vpjt)(\bar q'_p \tau^I \gamma^\mu q'_r)}$\\
$ $         &  &
$Q_{dB} $        & $(\bar q'_p \sigma^{\mu\nu} d'_r) \vp\, B_{\mu\nu}$ & 
${Q_{\vp u}}$            & ${(\vpj)(\bar u'_p \gamma^\mu u'_r)}$\\
$ $     &  &
$ $               &  &
${Q_{\vp d}}$            & ${(\vpj)(\bar d'_p \gamma^\mu d'_r)}$\\

%$$        &  &
%$ $   & \\
\hline \hline
%%%%%%%%%%%%%%%%%%%%%
 \multicolumn{2}{||c||}{} & \multicolumn{2}{|c||}{$\psi^4$} & \multicolumn{2}{|c||}{}\\
\hline
\multicolumn{2}{||c||}{} & $Q_{ll}$ & $(\bar l'_p \gamma_\mu
l'_r)(\bar l'_s \gamma^\mu l'_t)$ & \multicolumn{2}{|c||}{} \\
\hline\hline
\end{tabular}
\caption{\sl Dimension-6 operators contributing to $h\to Z\gamma$ decay. For
    brevity we suppress fermion chiral indices $L,R$. We follow here the
    notation of refs.~\cite{Grzadkowski:2010es,dedes:2017zog}. The operator
    class $\psi^{2}\phi^{2}D$ does not enter the $h\to \gamma\gamma$ amplitude.}
\label{tab:ops}
\end{table}
%%%%%%%%%%%%%%%%%%%%%%%%%%%%%%%

Let the lightest of the heavy-particle masses be of order $\Lambda$. Following
the decoupling theorem \cite{Appelquist:1974tg}, their effects at low energies
can be encoded in the renormalisation group running of the SM parameters in
addition to the appearance of local non-renormalisable operators. The later are
parameterised at low energies by a SMEFT Lagrangian,
which  takes the form
\begin{equation}
\mathcal{L} = \mathcal{L}_{\mathrm{SM}}^{(4)} + \sum_{X} C^{X} Q_X^{(6)}
+ \sum_{f} C^{\prime f} Q_f^{(6)} \,.
\label{Leff}
\end{equation}
Eq.~\eqref{Leff} contains the, renormalisable, SM Lagrangian
$\mathcal{L}_{\mathrm{SM}}^{(4)}$, the dimension-6 operators $Q_X^{(6)}$ that do
not involve fermion fields, and the dimension-6 operators $Q_f^{(6)}$ which are
operators that contain fermion fields.\footnote{The single $d=5$ lepton number
violating operator does not affect $h\to Z\gamma$ at one-loop.} All Wilson
coefficients should be rescaled by $\Lambda^2$, for instance $C^X \to
C^X/\Lambda^2$. We shall restore $1/\Lambda^2$ explicitly in
section~\ref{sec:results} later on. The primed coefficients $C^{\prime f}$ are
written in the gauge invariant Warsaw basis of ref.~\cite{Grzadkowski:2010es},
while the un-primed coefficients $C^{f}$ in fermion mass basis are defined in
ref.~\cite{dedes:2017zog}.

The operators contributing to the $h \to Z \gamma $ decay are collected in Table
\ref{tab:ops}. They are classified into 8 different classes according to the
notation of ref.~\cite{Grzadkowski:2010es}.	 There are in total 23 relevant
operators, not counting flavour structure and Hermitian conjugation. In unitary
gauge, the coefficient $C^{\varphi}$ associated with the operator $Q_{\varphi} =
(\varphi^\dagger \varphi)^3$ does \emph{not} appear in the calculation at
$\mathcal{O}(\Lambda^{-2})$ and therefore does not contribute in the final
amplitude.\footnote{On the contrary, in the $R_{\xi}$-gauges $C^{\varphi}$ 
enters in individual diagrams, but it cancels out completely in the final sum.
This adds to a list of several checks we performed in the final amplitude (cf.\
eq.~\eqref{eq:master2}).} The four-fermion operator $Q_{ll}$ enters indirectly
into the calculation through the relation between the vacuum expectation value
(VEV) and the Fermi coupling constant $G_{F}$. There are no contributions from
CP-violating operators up to $1/\Lambda^2$ terms in the EFT expansion. This is
based upon the fact that the SM amplitude is CP-invariant (symmetric in particle
momenta interchange) and all interference terms with CP-violating coefficients
(antisymmetric in particle momenta interchange) of $\mathcal{O}(1/\Lambda^2)$,
vanish identically. 

The 16 out of 23 operators affecting $h\to Z\gamma$ are identical with those
affecting the $h\to \gamma\gamma$ amplitude.\footnote{The operator
    $Q^{(3)}_{\varphi l}$ does in fact enters in the $h\to \gamma\gamma$
    amplitude, as well as in $h\to Z\gamma$, but only through the Fermi coupling
    constant redefinition, and not directly to  $h\to \gamma\gamma$ one-loop
amplitude.} The 7 operators that appear only in $h\to Z\gamma$ (those belonging
to category $\psi^2 \varphi^2 D$ of Table~\ref{tab:ops}) may provide assistance
in disentangling models for new physics in case of a $h\to Z\gamma$ experimental
discovery. This is interesting because, if perturbative decoupling of the UV
theory is assumed, the operators in $\psi^2 \varphi^2 D$ category are
potentially tree-level generated \cite{Arzt:1994gp,Einhorn:2013kja}. If the two
amplitudes, $h\to Z\gamma$ and $h\to \gamma\gamma$, are calculated in the same
renormalisation input scheme, we can compare the relative strengths of the
various contributions assuming dominance of one operator at a time. Within EFT
we should be able to pose predictions on possible sensitivity at LHC and future
colliders for the $h\to Z\gamma$ decay rate.

\section{Renormalisation of the $h\to Z\gamma$ Amplitude}
\label{sec:renormalisation}

Our renormalisation procedure follows an old but clear description invented by
A. Sirlin~\cite{Sirlin:1980nh}. This procedure has already been applied
successfully in a SMEFT calculation for $h\to \gamma\gamma$ in
ref.~\cite{Dedes:2018seb} and is quickly repeated  here  for completeness before
applying it to the calculation of the $h\to Z\gamma$ on-shell matrix element. 

%%%%%%%%%%%%%%%%%%%%%%
\subsection{Counterterms}
%%%%%%%%%%%%%%%%%%%%%%

We first start with the part of SMEFT Lagrangian bilinear in gauge fields in
gauge basis given in eq.~(3.14) of ref.~\cite{dedes:2017zog}, and write all bare
parameters as differences between renormalised parameters and corresponding
counterterms, for example $g_0 = g -\delta g$.
Then, mass diagonalisation for vector fields is performed by the matrix
$\mathbb{X}$ given in eq.~(3.19) of ref.~\cite{dedes:2017zog}. 
As we are only interested in an $S$-matrix element, we keep all fields
unrenormalised but multiplying the $h\to Z\gamma$ one-particle irreducible (1PI)
amplitude by proper LSZ constants~\cite{Lehmann:1954rq} for the external fields
of $h$, $\gamma$ and $Z$. In this way and after some algebra, counterterms are
generated and connected to self-energy corrections for vector bosons. We work at
one-loop in $\hbar$-expansion, and at $1/\Lambda^2$ in EFT expansion according
to our discussion below eq.~\eqref{Leff}.

The definition of 2- and 3-point 1PI correlation functions contains all
information we need to calculate the amplitude. Our definitions and conventions
follow directly those of refs.~\cite{Dedes:2018seb} and \cite{Sirlin:1980nh}. We
introduce the unrenormalised, but regularised, self-energies, that is 1PI
diagrams for scalars $s_{1,2}(=h)$, and vector bosons
$V_{1,2}(=W^\pm,Z,\gamma)$, 
%%%%%%%%%%%%%%%%%%%%%%%
\begin{align*}
\begin{gathered}
\begin{fmfgraph*}(70,40)
    \fmfleft{i}
    \fmfright{o}
    \fmf{dashes,label=$s_{1}$,l.side=right,label.dist=3}{v,i}    
    \fmf{dashes,label=$s_{2}$,l.side=left,label.dist=3}{v,o}
    \marrow{a}{down}{bot}{$q$}{i,v}
    \fmfblobbig{v}
\end{fmfgraph*}
\end{gathered}
\;&=-i\Pi_{s_{1}s_{2}}(q^{2})\,, \label{eq:hself} \numberthis\\
\begin{gathered}
\begin{fmfgraph*}(70,40)
    \fmfleft{i}
    \fmfright{o}
    \fmf{photon,label=$V_{1}^{\mu}$,l.side=left,label.dist=3}{i,v}   
    \fmf{photon,label=$V_{2}^{\nu}$,l.side=left,label.dist=3}{v,o}
    \marrow{a}{down}{bot}{$q$}{i,v}
    \fmfblobbig{v}
\end{fmfgraph*}
\end{gathered}
\;&=i\Pi^{\mu\nu}_{V_{1}V_{2}}(q^{2})
=iA_{V_{1}V_{2}}(q^{2}) g^{\mu\nu}
+iB_{V_{1}V_{2}}(q^{2}) q^{\mu}q^{\nu}\,, \label{eq:self} \numberthis \\
\begin{gathered}
\begin{fmfgraph*}(70,40)
    \fmfleft{i}
    \fmfright{o}
    \fmf{photon,label=$V_{1}^{\mu}$,l.side=left,label.dist=3}{i,v}   
    \fmf{photon,label=$V_{2}^{\nu}$,l.side=left,label.dist=3}{v,o}
    \marrow{a}{down}{bot}{$q$}{i,v}
    \fmfcounterterm{v}
\end{fmfgraph*}
\end{gathered}
\;&=i g^{\mu\nu} \delta m^{2}_{V_{1}V_{2}} 
+i q^{\mu}q^{\nu} \delta^{(q)} m^{2}_{V_{1}V_{2}} \,.\label{eq:cnt}  \numberthis
\end{align*}
%%%%%%%%%%%%%%%%%
We also include the definition \eqref{eq:cnt} for the vector boson counterterms
since these are needed in the final amplitude. Physical masses for vector
bosons, $M_{W}$ and $M_{Z}$, are defined to keep their tree-level form in SMEFT,
i.e.\ eqs.~(3.17) and (3.21) of ref.~\cite{dedes:2017zog}, by choosing the
corresponding counterterms such that
%%%%%%%%%%%%
\begin{equation}
\delta m_W^2 = \operatorname{Re} A_{WW}(M_W^2)\,, \quad \text{and} \quad 
\delta m_Z^2 = \operatorname{Re} A_{ZZ}(M_Z^2)\,.
\label{eq:dmv}
\end{equation}
%%%%%%%%%%%%%%
The physical masses $M_{W}$ and $M_{Z}$ for the $W^{\pm}$ and $Z$ vector bosons
are inputs in our calculation. In general, tadpole and tadpole-counterterm
diagrams also appear in the right side of \eqref{eq:dmv}. However,  one can
arrange  a  renormalisation condition where the tree-level VEV, $v$, is the
exact one up to one-loop order or beyond. Such a condition implies that tadpole
plus tadpole-counterterm diagrams vanish identically~\cite{Sirlin:1985ux}. In
addition, we define the weak mixing angle, $\theta_{W}$, through
\begin{equation}
c^2\equiv \cos^2\theta_W = \frac{M_W^2}{M_Z^2} \,, \qquad s^2\equiv 1- c^2 \,,
\qquad t\equiv \frac{s}{c} \,.
\label{eq:theta}
\end{equation}
%%%%%%%%%%%%%%%%%%%%%%%%%

%%%%%%%%%%%%%%%%%%%%%%%%%%%
\subsection{The Amplitude}
%%%%%%%%%%%%%%%%%%%%%%%%%%%

%%%%%%%%%%%%%%%%%%%%%%%
The on-shell $S$-matrix element for the $h\to Z\gamma$ amplitude can be written as
%%%%%%%%%%%%%5
\begin{equation}
%\braket{\gamma(\,\epsilon^\mu, p_1), Z(\epsilon^\nu, p_2) |S| h(q)} \ = \ 
%(2 \pi)^4 \delta^{(4)}(q-p_1-p_2) \: [ i \mathcal{A}^{\mu\nu}(h\to Z\gamma) ] \:
%\epsilon^*_{\mu} (p_1) \, \epsilon^*_{\nu}( p_2) \,,
\braket{\gamma(\,\epsilon^\mu, p_1), Z(\epsilon^\nu, p_2) |S| h(q)} = 
\sqrt{Z_h} \sqrt{Z_\gamma} \sqrt{Z_Z} \: [ i \mathcal{A}^{\mu\nu}(h\to Z\gamma) ] \:
\epsilon^*_{\mu} (p_1) \, \epsilon^*_{\nu}( p_2) \,,
\label{eq:s}
\end{equation}
%%%%%%%%%%%%
with $q=p_1+p_2$ the incoming Higgs boson momentum, and $p_1$ ($p_2$) the
outgoing four-momentum of photon ($Z$-boson) along with the polarisation
four-vectors, $\epsilon(p_1)$ [$\epsilon(p_2)$], respectively. Similar to the
mass counterterms $\delta m_V^2$ of \eqref{eq:dmv}, the LSZ factors $Z_h$,
$Z_\gamma$ and $Z_Z$ are calculated by the requirement for the full propagators
to look like those of free particle states asymptotically. Diagrammatically, the
amputated diagrams needed to sum up in eq.~\eqref{eq:s} are given in terms of 2-
and 3-point 1PI Feynman diagrams calculated on the mass shell, $p_1^2=0$, $p_2^2
= M_Z^2$ and $p_1\cdot p_2 = (M_h^2 - M_Z^2)/2$,
 %%%%%%%%%%%%
\begin{align*}
i \mathcal{A}^{\mu\nu}(h\to Z\gamma) \epsilon^*_{\mu} (p_1) \, \epsilon^*_{\nu}( p_2)
\ &= \quad
\begin{gathered}
\begin{fmfgraph*}(60,40)
    \fmfleft{i}
    \fmfright{o1,o2}
    \fmf{dashes,tension=1.3,label={\small $h$},label.dist=2}{v,i}
    \fmf{photon,label={\small $Z$},label.side=left,label.dist=4}{o1,v}
    \fmf{photon,label={\small $\gamma$},label.side=right,label.dist=4}{o2,v}
    \fmfsquare{v}
\end{fmfgraph*}
\end{gathered}
%%%%%%%%%%%%%%%%%
\quad+\quad
\\[2mm] &+\quad
%%%%%%%%%%%%%%%%%
\begin{gathered}
\begin{fmfgraph*}(60,40)
    \fmfleft{i}
    \fmfright{o1,o2}
    \fmf{dashes}{v,i}
    \fmf{photon}{o1,v}
    \fmf{photon}{o2,v}
    \fmfblobbig{v}
\end{fmfgraph*}
\end{gathered}
%%%%%%%%%%%%%%%%%
%%%%%%%%%%%%%%%%%
\quad+\quad
\begin{gathered}
\begin{fmfgraph*}(60,40)
    \fmfleft{i}
    \fmfright{o1,o2}
    \fmf{dashes,tension=1.6}{v,i}
    \fmf{phantom}{o1,v}
    \fmf{photon}{v,o2}
    \fmfsquare{v}
    \fmffreeze
    \fmf{photon,tension=1.5}{o1,vv}
    \fmf{photon,tension=1.2,label={\tiny $\gamma$},label.side=left,label.dist=3}{vv,v}
    \fmfblobsmall{vv}
\end{fmfgraph*}
\end{gathered}
\quad+\quad
%%%%%%%%%%%%%%%%
\begin{gathered}
\begin{fmfgraph*}(60,40)
    \fmfleft{i}
    \fmfright{o2,o1}
    \fmf{dashes,tension=1.6}{v,i}
    \fmf{phantom}{o1,v}
    \fmf{photon}{v,o2}
    \fmffreeze
    \fmf{photon,tension=1.5}{o1,vv}
    \fmf{photon,tension=1.2,label={\tiny $Z$},label.dist=3}{vv,v}
    \fmfblobsmall{vv}
\end{fmfgraph*}
\end{gathered}
%%%%%%%%%%%%%%
%%%%%%%%%%%%%%%%%
%\quad+\quad
\\[2mm] &+\quad
%%%%%%%%%%%%%%%%%
\begin{gathered}
\begin{fmfgraph*}(60,40)
    \fmfleft{i}
    \fmfright{o1,o2}
    \fmf{dashes,tension=1.6}{v,i}
    \fmf{photon}{o1,v}
    \fmf{photon}{v,o2}
    \fmfcounterterm{v}
\end{fmfgraph*}
\end{gathered}
%%%%%%%%%%%%%%%%
\quad+\quad
%%%%%%%%%%%%%%
\begin{gathered}
\begin{fmfgraph*}(60,40)
    \fmfleft{i}
    \fmfright{o1,o2}
    \fmf{dashes,tension=1.6}{v,i}
    \fmf{phantom}{o1,v}
    \fmf{photon}{v,o2}
    \fmfsquare{v}
    \fmffreeze
    \fmf{photon,tension=1.5}{o1,vv}
    \fmf{photon,tension=1.2,label={\tiny $\gamma$},label.side=left,label.dist=3}{vv,v}
    \fmfcounterterm{vv}
\end{fmfgraph*}
\end{gathered}
%%%%%%%%%%%%%%%%
\quad+\quad
%%%%%%%%%%%%%%
\begin{gathered}
\begin{fmfgraph*}(60,40)
    \fmfleft{i}
    \fmfright{o2,o1}
    \fmf{dashes,tension=1.6}{v,i}
    \fmf{phantom}{o1,v}
    \fmf{photon}{v,o2}
    \fmffreeze
    \fmf{photon,tension=1.5}{o1,vv}
    \fmf{photon,tension=1.2,label={\tiny $Z$},label.dist=3}{vv,v}
    \fmfcounterterm{vv}
\end{fmfgraph*}
\end{gathered}
\;.\numberthis
%%%%%%%%%%%%%%%%
\label{eq:diags}
\end{align*}
%%%%%%%%%%%%%%%% 
A square (``$\blacksquare$") in a vertex stands for a vertex generated by
\emph{only} $d=6$ operators. Shaded blobs in the second line denote, as before,
1PI 3-point $hZ\gamma$-vertex and 2-point $Z\gamma$- or $\gamma Z$-mixing at
one-loop, while diagrams with ``$\otimes$" symbol denote counterterms generated
following the procedure described above.

Before deriving the master formula for the $h\to Z\gamma$ decay amplitude, it is
worth noting a cancellation between some gauge non-invariant parts of the
counterterms. For this reason, let us focus on the third line of the diagrams in
\eqref{eq:diags} and collect the terms of the diagrams proportional to the
gauge invariant quantity
%%%%%%%%%%%%%%%%%%%%%%
\begin{equation}
\Delta^{\mu\nu} (p_1,p_2) = p_{1}^{\nu} \: p_{2}^{\mu} 
- (p_{1}\cdot p_{2}) g^{\mu\nu} \,.
\label{eq:dmn}
\end{equation}
Then, the gauge non-invariant leftovers are proportional to $g^{\mu\nu}$
(``pure-metric'' terms). For example, the $hZ\gamma$-vertex counterterm expands
diagramatically as, 
\begin{align}
    \begin{gathered}
    \begin{fmfgraph*}(60,40)
        \fmfleft{i}
        \fmfright{o1,o2}
        \fmf{dashes,tension=1.6}{v,i}
        \fmf{photon}{o1,v}
        \fmf{photon}{v,o2}
        \fmfcounterterm{v}
    \end{fmfgraph*}
    \end{gathered}
\quad\equiv\quad
    \begin{gathered}
    \begin{fmfgraph*}(60,40)
        \fmfleft{i}
        \fmfright{o1,o2}
        \fmf{dashes,tension=1.6}{v,i}
        \fmf{photon}{o1,v}
        \fmf{photon}{v,o2}
        \fmfv{decor.shape=circle,decor.filled=0,decor.size=4thick,label=$\boldsymbol{\times}$,label.dist=0}{v}
    \end{fmfgraph*}
    \end{gathered}
\Bigg{|}_{\Delta} \quad+\quad
    \begin{gathered}
    \begin{fmfgraph*}(60,40)
        \fmfleft{i}
        \fmfright{o1,o2}
        \fmf{dashes,tension=1.6}{v,i}
        \fmf{photon}{o1,v}
        \fmf{photon}{v,o2}
        \fmfv{decor.shape=circle,decor.filled=0,decor.size=4thick,label=$\boldsymbol{\times}$,label.dist=0}{v}
    \end{fmfgraph*}
    \end{gathered}
\Bigg{|}_{g} \quad ,
\end{align}
and similarly for the diagram containing the $Z\gamma$-mixing counterterm. We
can then prove that the sum of the ``pure-metric" contributions from the first
and the third diagram of \eqref{eq:diags} vanishes:\footnote{We remark here that
the counterterm for the $h \gamma\gamma$-vertex is gauge-invariant by
itself and, of course, zero in the SM.}
%%%%%%%%%%%%%%%
\begin{align}
    \begin{gathered}
    \begin{fmfgraph*}(60,40)
        \fmfleft{i}
        \fmfright{o1,o2}
        \fmf{dashes,tension=1.6}{v,i}
        \fmf{photon}{o1,v}
        \fmf{photon}{v,o2}
        \fmfv{decor.shape=circle,decor.filled=0,decor.size=4thick,label=$\boldsymbol{\times}$,label.dist=0}{v}
    \end{fmfgraph*}
    \end{gathered}
\Bigg{|}_{g} \quad+\quad
%%%%%%%%%%%%%%
\begin{gathered}
\begin{fmfgraph*}(60,40)
    \fmfleft{i}
    \fmfright{o2,o1}
    \fmf{dashes,tension=1.6}{v,i}
    \fmf{phantom}{o1,v}
    \fmf{photon}{v,o2}
    \fmffreeze
    \fmf{photon,tension=1.5}{o1,vv}
    \fmf{photon,tension=1.2,label={\tiny $Z$},label.dist=3}{vv,v}
    \fmfcounterterm{vv}
\end{fmfgraph*}
\end{gathered}
\Bigg{|}_{g} \quad=0\,.
\end{align}
%%%%%%%%%%%%%%%%%%
As a result, only the gauge invariant parts of these two counterterm diagrams
make it into the master formula for the amplitude below. Note that these
counterterm contributions exist even in the pure SM amplitude but usually not
discussed in the literature. One can of course exploit gauge invariance to start
with, as it was done for example in the first $h\to Z\gamma$ complete
calculation of ref.~\cite{Bergstrom:1985hp}, but it is really a nice cross-check
of the calculation to see how contributions turn out to be gauge-invariant,
respecting the usual Ward identities. Finally, note that the second diagram in
the third line of \eqref{eq:diags} is gauge invariant by itself.

We are now ready to present the on-shell reduced matrix element defined as
%%%%%%%%%%%%
\begin{equation}
\braket{\gamma(\,\epsilon^\mu, p_1), Z(\epsilon^\nu, p_2) |S| h(q)} =
(2\pi)^4 \delta^{(4)}(q-p_1-p_2)\: [ i \mathcal{M}^{\mu\nu}(h\to Z\gamma) ]\:
\epsilon_\mu^*(p_1) \epsilon_\nu^*(p_2) \,.
\label{eq:s2}
\end{equation}
%%%%%%%%%%%%%%
Adding the diagrams in \eqref{eq:diags} together and by comparing 
eqs.~\eqref{eq:s} and \eqref{eq:s2} we obtain
%%%%%%%%%%%%%%
\begin{align*}
\label{eq:master}
i \mathcal{M}^{\mu\nu}(h\to Z\gamma) 
&= 4i \, \Delta^{\mu\nu}(p_1,p_2) 
%\left[\, p_{1}^{\nu} p_{2}^{\mu} - (p_{1}\cdot p_{2}) g^{\mu\nu} \right]
\\ &\,\times
\Bigg{\{} -sc\, v\, C^{\varphi B} \left [ 1 + \mathcal{X}^{\varphi B} 
        - \frac{1}{t} \frac{ A_{Z\gamma}(M_{Z}^{2})+\delta m^{2}_{Z\gamma}}{M_{Z}^{2}}
        + t \frac{ A_{Z\gamma}(0)+\delta m^{2}_{Z\gamma}}{M_{Z}^{2}}
\right ] \\&\qquad
    + sc \, v \, C^{\varphi W} \left [ 1 + \mathcal{X}^{\varphi W} 
    + t \frac{ A_{Z\gamma}(M_{Z}^{2})+\delta m^{2}_{Z\gamma}}{M_{Z}^{2}}
    - \frac{1}{t} \frac{ A_{Z\gamma}(0)+\delta m^{2}_{Z\gamma}}{M_{Z}^{2}}
\right ] \\&\qquad
    +\frac{s^{2}-c^{2}}{2} \, v \, C^{\varphi WB} \left [ 1 + \mathcal{X}^{\varphi WB}
    - \frac{2sc}{s^{2}-c^{2}} \frac{ A_{Z\gamma}(M_{Z}^{2})+A_{Z\gamma}(0) 
    +2\delta m^{2}_{Z\gamma}}{M_{Z}^{2}}
\right ] \\&\qquad
    + \frac{1}{M_{W}} \overline{\Gamma}^{\mathrm{SM}} +
  \sum_{i \ne \varphi B, \varphi W, \varphi WB} \, v \, C^i \,
  \Gamma^i 
\Bigg{\}} \\&
  - 4i g^{\mu\nu} 
  \tfrac{1}{8}v(\bar{g}^{2}+\bar{g}^{\prime 2})
  \left[ 1+v^{2}C^{\varphi \square}+\tfrac{3}{4} v^{2}C^{\varphi D} 
    + 2sc\, v^{2} C^{\varphi WB} \right]
\frac{A_{Z\gamma}(0)}{M_{Z}^{2}}
\\&
-4i (p_{1}\cdot p_{2}) g^{\mu\nu}\left[ \frac{1}{M_{W}}
    \overline{\Gamma}_{g}^{\mathrm{SM}} +
  \sum_{i} \, v C^i \,
  \Gamma_{g}^{i} \right]
\,. \numberthis
\end{align*}
%%%%%%%%%%%%%%%%
Eq.~\eqref{eq:master} is the master formula for the $h\to Z\gamma$ on-shell
amplitude. The gauge-invariant quantity $\Delta^{\mu\nu}(p_1,p_2)$ has been
defined in \eqref{eq:dmn} while the self-energies and the counterterm $\delta
m^2_{Z\gamma}$ in eqs.~\eqref{eq:self} and \eqref{eq:cnt}, respectively.
Moreover, in \eqref{eq:master} and for brevity, we defined the quantity
%%%%%%%%%%%%%%%
\begin{equation}
\mathcal{X}^{i}\equiv 
\Gamma^{i}-\frac{\delta C^{i}}{C^{i}}-\frac{\delta v}{v}
+\tfrac{1}{2}\Pi_{hh}^\prime(M_{h}^{2})
+\tfrac{1}{2}A^{\prime}_{ZZ}(M_{Z}^{2})
+\tfrac{1}{2}A^{\prime}_{\gamma\gamma}(0)\,,
\label{eq:Xcal}
\end{equation}
%%%%%%%%%%%%%%%%
where $i=\varphi B,\varphi W,\varphi WB$. In \eqref{eq:Xcal}, $\Gamma^i$ stands
for 1PI contributions from the first diagram in the 2nd line of
\eqref{eq:diags}. $\overline{\Gamma}^{\mathrm{SM}}$ is the SM contribution from
triangle diagrams with $W$-bosons and fermions. In addition, $\delta C^i$ and
$\delta v$ are counterterms for the Wilson coefficients with $i=\varphi
B,\varphi W,\varphi WB$ and the VEV, respectively. The $\delta v/v$ counterterm
is specified in eqs.~(3.18)--(3.20) of ref.~\cite{Dedes:2018seb} after following
the renormalisation scheme of refs.~\cite{Sirlin:1980nh,Sirlin:1985ux}. The
coefficients $C^i$ (and in fact all Wilson coefficients 
in \eqref{Leff}) can be readily transformed in $\overline{\mathrm{MS}}$-scheme,
$C -\delta C \to C(\mu) - \delta C$. As usual, in this scheme the counterterms
$\delta C^i$ subtract infinite parts proportional to $(\frac{2}{4-d}-\gamma+\log
4 \pi )$ and can be read directly from eqs.~(3.23)--(3.25) of
ref.~\cite{Dedes:2018seb} as they have been adapted from
refs.~\cite{Jenkins:2013zja, Jenkins:2013wua, Alonso:2013hga}.
We confirm, even analytically, that these counterterms are capable of
subtracting all infinities arising from the one-loop diagrams.
The last three terms in the right-hand side arise from the product of the square
roots of the LSZ factors in \eqref{eq:s} where the prime denotes derivative with
respect to $q^2$, for example $\Pi_{hh}^\prime(q^2)= d \Pi_{hh}(q^2)/d q^2$.
Finally, note that the hadronic contributions from light quarks in
$A_{\gamma\gamma}^{\prime}(0)$, as given in eq.~(4.21) of
ref.~\cite{Dedes:2018seb}, have been taken into account since they have an
important contribution (one order of magnitude) in the non-logarithmic parts of
the one-loop amplitude.

Note that eq.~\eqref{eq:master} is divided in two parts: the first part is
proportional to the gauge-invariant quantity $\Delta^{\mu\nu}$, while the second
part (last two lines of eq.~\eqref{eq:master}) is proportional to $g^{\mu\nu}$
and, therefore, is \emph{not} gauge-invariant and violates the Ward-identity for
charge conservation. We have proved that for every gauge-fixing choice these
contributions vanish. 
To be more specific, we have checked explicitly that in unitary gauge
$A_{Z\gamma}(0)=0$ and that there are no leftover corrections proportional to
$g^{\mu\nu}$, i.e.\ $\overline{\Gamma}_{g}^{\mathrm{SM}}=\Gamma_g^i=0$. What
happens in $R_\xi$-gauges is discussed at the end of subsection~\ref{sub:xi}.

We are now ready to write the $h\to Z\gamma$ amplitude at one-loop and at
$1/\Lambda^2$ in EFT expansion. After removing the last two lines in
\eqref{eq:master} and checking that infinities cancel when applying the
counterterms $\delta C^i$, we arrive at the matrix element  
%%%%%%%%%%%%%%%%%
\begin{align*}
\label{eq:master2}
i \mathcal{M}^{\mu\nu}(h\to Z\gamma) 
&= 4i \, \Delta^{\mu\nu}(p_1,p_2) 
%\left[\, p_{1}^{\nu} p_{2}^{\mu} - (p_{1}\cdot p_{2}) g^{\mu\nu} \right]
\\ &\,\times
\Bigg{\{} -sc\, v\, C^{\varphi B} \left [ 1 + \mathcal{X}^{\varphi B} 
        - \frac{1}{t} \frac{ A_{Z\gamma}(M_{Z}^{2})+\delta m^{2}_{Z\gamma}}{M_{Z}^{2}}
        + t \frac{ A_{Z\gamma}(0)+\delta m^{2}_{Z\gamma}}{M_{Z}^{2}}
\right ] \\&\qquad
    + sc \, v \, C^{\varphi W} \left [ 1 + \mathcal{X}^{\varphi W} 
    + t \frac{ A_{Z\gamma}(M_{Z}^{2})+\delta m^{2}_{Z\gamma}}{M_{Z}^{2}}
    - \frac{1}{t} \frac{ A_{Z\gamma}(0)+\delta m^{2}_{Z\gamma}}{M_{Z}^{2}}
\right ] \\&\qquad
    +\frac{s^{2}-c^{2}}{2} \, v \, C^{\varphi WB} \left [ 1 + \mathcal{X}^{\varphi WB}
    - \frac{2sc}{s^{2}-c^{2}} \frac{ A_{Z\gamma}(M_{Z}^{2})+A_{Z\gamma}(0)
    +2\delta m^{2}_{Z\gamma}}{M_{Z}^{2}}
\right ] \\&\qquad
    + \frac{1}{M_{W}} \overline{\Gamma}^{\mathrm{SM}} +
  \sum_{i \ne \varphi B, \varphi W, \varphi WB} \, v \, C^i \,
  \Gamma^i 
\Bigg{\}}_{\mathrm{finite} }
\,, \numberthis
\end{align*}
%%%%%%%%%%%%%%%%
which is gauge invariant and renormalisation scale $\mu$-independent, in a sense
that $ \mu \,d\mathcal{M}^{\mu\nu}/d\mu = 0$. The subscript ``finite" means that
infinities proportional to $(\frac{2}{4-d}-\gamma+\log 4\pi )$ have been removed
from expressions such as $A_{VV}$, $A_{VV}^\prime$, $\Gamma^i$, etc, with
counterterms $\delta C^i$ removed from the quantity $\mathcal{X}^i$ of
\eqref{eq:Xcal} as well. All self-energies but $A_{Z\gamma}(M_Z^2)$ and
$A^{\prime}_{ZZ}(M_{Z}^2)$ appearing in \eqref{eq:master2} are given
analytically in general $R_\xi$-gauges, in Appendix A of
ref.~\cite{Dedes:2018seb} (see also \cite{Marciano:1980pb} for formulae in
$\xi=1$).
It is obvious from \eqref{eq:master2} that self-energies for the Higgs or
vector bosons should be calculated \emph{only} in the SM, not (necessarily) in
SMEFT. The three-point vertex functions $\Gamma^i$ are in general too lengthy
and is not really illuminating to be given here.

Although we leave the expression \eqref{eq:master2} for the matrix element in a
slightly involved form, it can be reduced further by noting the following. As in
the case of the $h\to \gamma\gamma$ amplitude, there is a remarkable relation
between factors multiplying the coefficients $C^{\varphi B}$ and $C^{\varphi W}$
when replacing 
 \begin{equation}
 \tan\theta_W \to - \frac{1}{\tan\theta_W} \,,
 \label{eq:t}
\end{equation} 
while on the other hand, factors multiplying $C^{\varphi WB}$ in
\eqref{eq:master2} remain invariant.
In addition, elementary trigonometric relations may reduce
eq.~\eqref{eq:master2} further. For example, by using
$\tan\theta_W-1/\tan\theta_W = -2 \cot^2 (2\theta_W)$ one may factor out $\delta
m^2_{Z\gamma}/M_Z^2$ terms. We believe, however, that eq.~\eqref{eq:master2} is
more transparent and easily understood when read in conjunction with the list of
diagrams and counterterms of eq.~\eqref{eq:diags}. 

Finally, some words about calculating the diagrams appearing in the shaded blobs
of \eqref{eq:diags}. We used the Feynman Rules of ref.~\cite{dedes:2017zog},
given in general $R_\xi$-gauges, and passed them manually to the
\texttt{Mathematica} package \texttt{FeynCalc}
\cite{Mertig:1990an,Shtabovenko:2016sxi}. The Feynman integrals are regulated
with dimensional regularisation \cite{tHooft:1971qjg} with the Dirac algebra
performed in $d$-dimensions. The result is reduced to basic Passarino-Veltman
functions~\cite{Passarino:1978jh}. We then checked expressions for analytic
functions, some of them presented in ref.~\cite{Dedes:2018seb}, against the
numerical library \texttt{LoopTools} \cite{Hahn:1998yk,vanOldenborgh:1989wn}.
The most crucial (and time consuming) test is the gauge-fixing parameter
independence of the amplitude \eqref{eq:master}.

%%%%%%%%%%%%%%%%%%%%%%%%%%%%%%%%
\subsection{Gauge-fixing parameters cancellation}
\label{sub:xi}
%%%%%%%%%%%%%%%%%%%%%%%%%%%%%%%%

Since the cancellation in the amplitude of the gauge-fixing parameters,
collectively denoted as $\xi$, is a very involved and important cross-check of
the validity of our calculation, let us give here some insight on this
particular computational task. In general, there are two different ways of how
$\xi$-dependent contributions arise in SMEFT. Let us call the result one finds
by subtracting the unitary gauge result from the full result in
$\mathrm{R}_{\xi}$-gauges the $\xi$-dependent result. For an operator $C^{i}$
there are \emph{explicit} $\xi$-dependent contributions, coming from the
$\xi$-dependent result which is proportional to the Wilson coefficient $C^{i}$.
There are also \emph{implicit} contributions, coming from the $\xi$-dependent
SM-like result by Taylor-expanding the masses with $C^{i}$ as an expansion
parameter.

In the $h\to Z\gamma$ process there are two independent gauge-fixing parameters,
$\xi_{W}$ and $\xi_{Z}$. We therefore prove the $\xi$-cancellation in the
amplitude for each of these parameters \emph{independently}. Interchanging
between gauge-fixing parameters is a great advantage of the Feynman rules
written in general $R_\xi$-gauges in ref.~\cite{dedes:2017zog}. We also checked
gauge-invariance without any renormalisation scheme. In this case, one has to
add a Higgs tadpole diagram in the ``hhAZ'' vertex. 
As explained in eq.~\eqref{eq:master}, the last two lines do not appear in the
unitary gauge at all. On the other hand, each of the terms in these lines
contributes in the $\xi$-dependent part, so one has to prove that they add to
zero. Note that there are explicit contribution from the SMEFT $\Gamma$s in the
last line as well as from the vertex and the $Z\gamma$-mixing in the penultimate
line, and also implicit contributions from the $Z$-boson mass and the
$Z\gamma$-mixing in the penultimate line and the $\xi$-dependent SM result in
the last line.

It is important to stress here the analytic result of the $Z\gamma$-mixing in
SMEFT with $d=6$ operators. One can prove that the result is simply given by
\begin{equation}
\label{eq:za-mixing}
A_{Z\gamma}^{\mathrm{SMEFT}}(0) = \left(1+\tfrac{1}{2}v^{2} C^{\varphi D}\right) 
A_{Z\gamma}^{\mathrm{SM}}(0) \,,
\end{equation}
where $A_{Z\gamma}^{\mathrm{SM}}(0)$ is the SM-like value at $q^2=0$, given
analytically in eq.~(A.6) of ref.~\cite{Dedes:2018seb}. Note that
$A_{Z\gamma}^{\mathrm{SM}}(0)$ is a function of the SMEFT couplings, the VEV and
the $W$-boson mass. Therefore, the SM-like and the SM values coincide. We
believe that \eqref{eq:za-mixing} has interesting consequences in the general
SMEFT renormalisation program.

Each coefficient has its own unique way of how the $\xi$-cancellation occurs. As
an example, let us discuss here the $C^{\varphi WB}$ coefficient. Since
$A_{Z\gamma}^{\mathrm{SMEFT}}$ doesn't depend on this coefficient (either
explicitly or implicitly) and the vertex contribution cancels that of the
$Z$-boson mass, $C^{\varphi WB}$ cancels trivially in the penultimate line.
Therefore, the implicit and explicit contributions from the last line should
cancel among each other, which we have proved that this is exactly the case.

%%%%%%%%%%%%%%%%%%%%
\section{Results}
\label{sec:results}
%%%%%%%%%%%%%%%%%%%%

\subsection{$h\to Z\gamma$ in the Standard Model and the input parameters scheme}

As it is well known, the $h\to Z\gamma$ SM contribution,
$\overline{\Gamma}^{\mathrm{SM}}/M_{W}$, in eq.~\eqref{eq:master2} is a sum of
one-loop diagrams with only $W^\pm$ bosons and  charged fermions, $f$,
circulating in the loop. In terms of the SMEFT parameters $\{ \bar{g},
\bar{g}^\prime, v \}$, defined in ref.~\cite{dedes:2017zog}, we find that
\begin{equation}
\frac{\overline{\Gamma}^{\mathrm{SM}}}{M_{W}} =
\frac{\bar{g}\bar{g}^{\prime}}{16 \pi^{2} v} 
\left[ \sum_{f} N_{c,f} Q_{f} \left(T^{3}_{f}-2 Q_{f} \frac{ \bar{g}^{\prime
2}}{\bar{g}^{2}+\bar{g}^{\prime 2}}\right) I_{f} + \frac{\bar{g}^{2}}{2
(\bar{g}^{2}+\bar{g}^{\prime 2})} I_{W} \right] \,,
\label{smlikeresult}
\end{equation}
where the electromagnetic fermion charges and the third component of the weak
isospin are given by
%%%%%%%%%%%%%%%%%%%%  
\begin{equation}
Q_{f}=
\begin{cases}
0, &\text{for} \quad f=\nu_{e},\nu_{\mu},\nu_{\tau}\\
-1, &\text{for} \quad f=e,\mu,\tau\\
2/3, &\text{for} \quad f=u,c,t\\
-1/3, &\text{for} \quad f=d,s,b\\
\end{cases} \quad \text{and} \quad
T_{f}^{3}=
\begin{cases}
1/2, &\text{for} \quad f=\nu_{e},\nu_{\mu},\nu_{\tau},u,c,t\\
-1/2, &\text{for} \quad f=e,\mu,\tau,d,s,b
\end{cases}\,.
\label{eq:charges}
\end{equation}
%%%%%%%%%%%%%%%%%%%%%%%%%%%
and the colour factor $N_{c,f}$ is equal to $1$ for leptons and $3$ for quarks.
In \eqref{smlikeresult} $I_{f}$ and $I_{W}$ contain the contribution from the
fermionic and the bosonic sector respectively. Explicitly, these quantities are
given in terms of PV functions as:\footnote{Our notation for PV-functions is
identical to those of {\tt LoopTools} in ref.~\cite{Hahn:1998yk}.}
\begin{align}
I_{f}= \frac{m^{2}_{f}}{(M^{2}_{h}-M^{2}_{Z})^{2}} &\Bigg\{2
    M^{2}_{Z}\left[ B_{0}(M^{2}_{h},m^{2}_{f},m^{2}_{f}) -
    B_{0}(M^{2}_{Z},m^{2}_{f},m^{2}_{f}) \right] 
\nonumber \\&\quad
-(M^{2}_{h} -M^{2}_{Z}) \left[ (M^{2}_{h} -M^{2}_{Z} -4 m^{2}_{f})
C_{0}(0,M^{2}_{h},M^{2}_{Z},m^{2}_{f},m^{2}_{f},m^{2}_{f}) -2\right] \Bigg\}\,,
\end{align}
and
\begin{align}
I_{W} &= \frac{1}{(M^{2}_{h}-M^{2}_{Z})^{2}} 
\Bigg\{ \frac{M^{2}_{Z}}{M^{2}_{W}} \left[ M^{2}_{h} (M^{2}_{Z}-2M^{2}_{W}) +
    2M^{2}_{W} (M^{2}_{Z}-6 M^{2}_{W} ) \right]
\nonumber \\ & \qquad
\times \left[ B_{0}(M^{2}_{h},M^{2}_{W},M^{2}_{W}) -
B_{0}(M^{2}_{Z},M^{2}_{W},M^{2}_{W}) \right]
\nonumber \\ \qquad&
+ \frac{M^{2}_{h}-M^{2}_{Z}}{M^{2}_{W}}  \Big[ 2
    M^{2}_{W} \left( M^{2}_{h}(6 M^{2}_{W}-M^{2}_{Z})-12 M^{4}_{W}-6 M^{2}_{W}
    M^{2}_{Z}+2 M^{4}_{Z}\right)
\nonumber \\ & \qquad
\times C_{0}(0,M^{2}_{h},M^{2}_{Z},M^{2}_{W},M^{2}_{W},M^{2}_{W}) +M^{2}_{h}
(M^{2}_{Z} -2 M^{2}_{W}) +2 M^{2}_{W} (M^{2}_{Z} -6 M^{2}_{W}) \Big] \Bigg\}\,.
\end{align}
%%%%%%%%%%%%%%%%%
We have proved explicitly that the SM matrix element is finite, gauge invariant
and gauge-fixing parameter independent. 

We can express the SM-like result of eq.~\eqref{smlikeresult}, or for that
matter any other contribution in \eqref{eq:master2}, in terms of well-measured
quantities that will be taken as inputs in evaluating the $h\to Z\gamma$
amplitude. The set of well-measured quantities we have chosen, contains
$G_{F}$, $M_{W}$ and $M_{Z}$. Using the following expressions for
$\bar{g}^\prime,\bar{g}$ and $v$ as functions of the Fermi coupling constant
$G_{F}$, the physical $W$-boson mass, $M_{W}$, and the physical $Z$-boson mass,
$M_{Z}$, we obtain:
%%%%%%%%%%%%%%%%%%%%%%%%%%%%%%%%%%%%
\begin{align}
\bar{g}^{\prime} = 2^{5/4} \sqrt{M^{2}_{Z}-M^{2}_{W}} \sqrt{G_{F}} \Bigg[ 1 & -
    \frac{1}{2\sqrt{2} G_{F}} \left( \frac{C^{\varphi l(3)}_{11}}{\Lambda^{2}} +
\frac{C^{\varphi l(3)}_{22}}{\Lambda^{2}} - \frac{C^{ll}_{1221}}{\Lambda^{2}}
\right) \nonumber \\&
- \frac{M^{2}_{Z}}{4\sqrt{2} G_{F}(M^{2}_{Z}-M^{2}_{W})} \left( \frac{C^{\varphi
D}}{\Lambda^{2}} + 4 \frac{M_{W}}{M_{Z}} \sqrt{1 - \frac{M^{2}_{W}}{M^{2}_{Z}}}
\frac{C^{\varphi W B}}{\Lambda^{2}}  \right) 
\Bigg]\,,
\label{eq:gp}
\end{align}
%%%%%%%%%%%%%%%%%%%%%%%%%%%%%%%%%%%%%%%
\begin{equation}
\bar{g} = 2^{5/4} M_{W} \sqrt{G_{F}} \left[ 1 - \frac{1}{2\sqrt{2}G_{F}} \left(
\frac{C^{\varphi l(3)}_{11}}{\Lambda^{2}} + \frac{C^{\varphi
l(3)}_{22}}{\Lambda^{2}} - \frac{C^{ll}_{1221}}{\Lambda^{2}} \right)
\right]\,,
\label{eq:g}
\end{equation}
%%%%%%%%%%%%%%%%%%%%%%%%%%%%%
\begin{equation}
v = \frac{1}{2^{1/4} \sqrt{G_{F}}} \left[ 1 + \frac{1}{2\sqrt{2}G_{F}} \left(
\frac{C^{\varphi l(3)}_{11}}{\Lambda^{2}} + \frac{C^{\varphi
l(3)}_{22}}{\Lambda^{2}} - \frac{C^{ll}_{1221}}{\Lambda^{2}} \right)
\right] \,.
\label{eq:v}
\end{equation}
%%%%%%%%%%%%%%%%%%%%%%%%%%%%%%%%%%%%

Finally we can express  the parameters in eq.~\eqref{smlikeresult} as a function
of the experimental quantities $G_{F}$, $M_{W}$ and $M_{Z}$ taken from
PDG~\cite{Tanabashi:2018oca}. The reason for choosing the input scheme
$\{G_F,M_W,M_Z\}$ is twofold: first, it has natural implementation\footnote{At
least more natural than the scheme with the input set $\{ \alpha_{em},G_F,M_Z
\}$.} into our renormalisation prescription discussed already in
section~\ref{sec:renormalisation} and especially into the simple definition of
the weak mixing angle in eq.~\eqref{eq:theta}\footnote{Note that the expression
    for $\bar{g}^\prime$ in  eq.~\eqref{eq:gp} becomes much simpler upon
    substitution of the weak mixing angle definition of \eqref{eq:theta}. Then
    the second line of \eqref{eq:gp} reads:
\begin{equation*}
- \frac{1}{4\sqrt{2}\, s^{2} G_{F}} \left( \frac{C^{\varphi D}}{\Lambda^{2}} + 4
sc \frac{C^{\varphi W B}}{\Lambda^{2}}\right)\;.
\end{equation*}}
and second it is a scheme that is becoming increasingly popular after
refs.~\cite{Dawson:2018liq,Dawson:2018pyl} with whom we would like to compare
our results. Other advantages of this scheme have also been put forward by
ref.~\cite{Brivio:2017bnu}.

After replacing $\bar{g}^\prime$, $\bar{g}$ and $v$ in eq.~\eqref{smlikeresult} 
with eqs.~\eqref{eq:gp}--\eqref{eq:v}, it is rather more instructive to 
present also the numerical result here. This reads
%%%%%%%%%%%%%%%%%%%%%%%%%%%%%%%%%%%%%%%
\begin{align}
\frac{\overline{\Gamma}^{\mathrm{SM}}}{M_{W}}%|_{(h \to Z \gamma)}
=& -(1.43 \times 10^{-5}-1.11 \times 10^{-8} i)  \nonumber \\ & +(1.07+1.38 \times 10^{-4} i)
\frac{C^{\varphi W B}}{\Lambda^{2}}
+(0.64 + 8.28 \times 10^{-5} i) \frac{C^{\varphi D}}{\Lambda^{2}}
\nonumber \\ &
+(1.30 - 1.00 \times 10^{-3} i) \left[\frac{C^{\varphi l
(3)}_{11}}{\Lambda^{2}}+\frac{C^{\varphi l
(3)}_{22}}{\Lambda^{2}}-\frac{C^{ll}_{1221}}{\Lambda^{2}}\right].
\label{eq:numhZg}
\end{align}
%%%%%%%%%%%%%%%%%%%%%%%%%%%%%%%%%%%%%%%%
As one can see, the imaginary part of the SM-like amplitude is more than three
orders of magnitude smaller than the real part and can be safely ignored in the
following. Our result agrees with ref.~\cite{Gunion:1989we} and partially with
refs.~\cite{Bergstrom:1985hp,Djouadi:2005gi}.\footnote{We have a minus sign
    difference in the term before the last parenthesis of eq.~(4) of
ref.~\cite{Bergstrom:1985hp}. Furthermore, our SM result agrees with
ref.~\cite{Djouadi:2005gi} only if the branches of the piecewise function
$g(\tau)$ in eq.~(2.56) are reversed.}
The pure SM contribution, $\overline{\Gamma}^{\mathrm{SM}}/M_{W}$, can be
factored out in the amplitude of \eqref{eq:master2} and after squaring and
integrating over the phase space of the final state particles, $\gamma$ and $Z$,
one can easily find the decay rate for $h\to Z\gamma$ in the SM and in SMEFT. It
is then useful to express our results in terms of the quantity 
\begin{equation}
\mathcal{R}_{h \to Z \gamma}=\frac{\Gamma(\mathrm{SMEFT},h \to Z
\gamma)}{\Gamma(\mathrm{SM},h \to Z \gamma)}\equiv 1+ \delta \mathcal{R}_{h \to Z \gamma},
\end{equation}
and compare with the experimental bound of eq.~\eqref{eq:RhZg}. In the next
subsection we present corrections for $\delta \mathcal{R}_{h \to Z \gamma}$ from
new physics in the form of running Wilson coefficients of the operators listed
in Table \ref{tab:ops}. In addition, we search for correlations with an
analogous expression arising from the $h\to \gamma\gamma$ decay.

%%%%%%%%%%%%%%%%%%%%%
\subsection{Semi-numerical expression for the ratio $\mathcal{R}_{h\to Z\gamma}$}
%%%%%%%%%%%%%%%%%%%%%%

In this section we finally present our results for $\delta \mathcal{R}_{h \to Z
\gamma}$. As in ref.~\cite{Dedes:2018seb}, we shall separate constant and
renormalisation scale $\mu$-dependent logarithmic parts which multiply RGE
running Wilson coefficients, $C(\mu)$. In ``Warsaw" mass-basis of
ref.~\cite{dedes:2017zog}, by exploiting the input parameters scheme $\{ G_F,
M_W, M_Z \}$ with the new-physics scale $\Lambda$ written in TeV-units, we
find:\footnote{Our result is in agreement with the revised (arXiv v3)
    version of ref.~\cite{Dawson:2018pyl}.}
%%%%%%%%%%%%%%%%%%%%%%%
\begin{align}
\label{eq:num2}
\delta \mathcal{R}_{h\to Z\gamma} & \simeq
 0.18 \frac{ C^{ll}_{1221} - C^{\varphi l(3)}_{11} - 
C^{\varphi l(3)}_{22}}{\Lambda^2} +
0.12  \frac{C^{\varphi \Box} - C^{\varphi D} }{\Lambda^2}  
\nonumber \\[2mm] &\quad
- 0.01  \frac{C^{d\varphi }_{33} - C^{u \varphi }_{33}}{\Lambda^2} 
+ { 0.02 \frac{ C^{\varphi u}_{33} + C^{\varphi q(1)}_{33} - 
C^{\varphi q(3)}_{33}}{\Lambda^2} }
\nonumber \\[2mm]&\quad
+ \left[14.99 - 0.35 \log \frac{\mu^{2}}{M^{2}_{W}}\right]\frac{C^{\varphi
B}}{\Lambda^{2}}
- \left[14.88 - 0.15 \log \frac{\mu^{2}}{M^{2}_{W}}\right]\frac{C^{\varphi
W}}{\Lambda^{2}} \nonumber \\[2mm] &\quad
+ \left[9.44 - 0.26 \log \frac{\mu^{2}}{M^{2}_{W}}\right]\frac{C^{\varphi
WB}}{\Lambda^{2}}
+ \left [ 0.10 - 0.20 \log
\frac{\mu^{2}}{M^{2}_{W}}\right]\frac{C^{W}}{\Lambda^{2}} \nonumber \\[2mm] &\quad
- \left [ 0.11 - 0.04 \log
\frac{\mu^{2}}{M_W^2}\right]\frac{C^{uB}_{33}}{\Lambda^{2}} 
+ \left [ 0.71 -
0.28 \log \frac{\mu^{2}}{M_W^2}\right]\frac{C^{uW}_{33}}{\Lambda^{2}} \nonumber
\\[2mm] &\quad
- \left [ 0.01 +
0.00 \log \frac{\mu^{2}}{M_W^2}\right]\frac{C^{uW}_{22}}{\Lambda^{2}}
- \left [ 0.01 +
0.00 \log \frac{\mu^{2}}{M_W^2}\right]\frac{C^{dW}_{33}}{\Lambda^{2}}
+ \ldots \,,
\end{align}
%%%%%%%%%%%%%%%%%%%%%%%
where the ellipses denote contributions from operators that are less than $0.01
\times C/\Lambda^2$. Note that the VEV appearing at tree-level introduces
one-loop corrections when exchanged for the Fermi constant through
$\overline{G}_F = 1 / [\sqrt{2} v^2 (1 - \Delta r)]$. We follow here the same
procedure as in ref.~\cite{Dedes:2018seb} below eq.~(4.16). Formula
\eqref{eq:num2} should be renormalisation scale ($\mu$) independent at one-loop
and up to terms with $1/\Lambda^2$ in EFT expansion.
Assuming that Higgs boson production is not affected by the operators listed in
Table \ref{tab:ops}, the current experimental bound of \eqref{mu} sets rather
weak constraints on tree-level SMEFT Wilson coefficients. As an example, for
$\mu=M_W$ we obtain
%%%%%%%%%%%%%%
\begin{equation}
\frac{|C^{\varphi B}|}{\Lambda^2} \lesssim \frac{0.4}{(1\, \mathrm{TeV})^2} \,, \qquad 
\frac{|C^{\varphi W}|}{\Lambda^2} \lesssim \frac{0.4}{(1\, \mathrm{TeV})^2} \,, \qquad
\frac{|C^{\varphi W B}|}{\Lambda^2} \lesssim \frac{0.7}{(1\, \mathrm{TeV})^2} \,.
\end{equation}
%%%%%%%%%%%%%%%
For loop-induced operators, the logarithmic part is of the same order of
magnitude as of the constant part. Contributions in the first and second line
of \eqref{eq:num2} arise from finite fermionic triangle diagrams that just
rescale the SM result. Wilson coefficients $C^{\varphi u}_{33}$,
$C^{\varphi q (1)}_{33}$, $C^{\varphi q (3)}_{33}$ are the new operators
appearing now in $h\to Z\gamma$ decay relative to $h\to \gamma\gamma$ (see
Table~\ref{tab:ops}). Interestingly, out of many operators only three made a
contribution for more than 1\%  and in fact they are just barely pass that
threshold!\footnote{We consider 1\% of corrections as an indicative limit that
LHC can reach for $\delta \mathcal{R}_{h\to Z\gamma}$ at later stages of its
run.} 

How to use eq.~\eqref{eq:num2}? First, decouple heavy particles  from a more
fundamental theory. Match to Warsaw-basis operators relevant for $h\to Z\gamma$,
listed in Table~\ref{tab:ops}. Set the coefficients, $C(\mu)$ at a scale
$\mu=\Lambda$. Use RGEs to run the parameters down to the Higgs mass scale ---
one could use dedicated codes for this purpose like those in
refs.~\cite{Celis:2017hod,Aebischer:2018bkb}. Plug in the results for
$C(\mu=M_h)$ coefficients in eq.~\eqref{eq:num2} and obtain $\delta
\mathcal{R}_{h\to Z\gamma}$. As long as discussing the same physical process in
the same input parameter scheme, the result should be unambiguous.

Keeping in mind the current experimental sensitivity for $h\to Z\gamma$,
eq.~\eqref{eq:num2} is not of much use. It is however, quite interesting to
check for a $h\to Z\gamma$ projective reach  by comparing $\delta
\mathcal{R}_{h\to Z\gamma}$ of eq.~\eqref{eq:num2} with $\delta
\mathcal{R}_{h\to \gamma\gamma}$ taken from ref.~\cite{Dedes:2018seb} but
translated into the $\{ G_F, M_W, M_Z \}$ input scheme. We have, 
%%%%%%%%%%%%%%%%
\begin{align}
\label{eq:h2gg-num} 
    \delta \mathcal{R}_{h\to \gamma\gamma} &
\simeq 0.18 
\frac{C^{ll}_{1221} - C^{\varphi l (3)}_{11} - C^{ \varphi l (3)}_{22}}
{\Lambda^2}  + 0.12 \frac{C^{\varphi \Box} - 2 C^{\varphi D} }{\Lambda^2}
\nonumber \\[2mm]&\quad
- 0.01  \frac{C^{e\varphi}_{22} + 4 C^{e\varphi}_{33} + 5
   C^{u\varphi }_{22} + 2 C^{d\varphi }_{33} - 3 C^{u \varphi
  }_{33}}{\Lambda^2}  \nonumber \\[2mm]&\quad
- \left[48.04 - 1.07 \log
  \frac{\mu^{2}}{M^{2}_{W}}\right]\frac{C^{\varphi
    B}}{\Lambda^{2}}
- \left[14.29 - 0.12 \log
  \frac{\mu^{2}}{M^{2}_{W}}\right]\frac{C^{\varphi
  W}}{\Lambda^{2}} \nonumber \\[2mm]&\quad
+ \left[26.17 - 0.52 \log
  \frac{\mu^{2}}{M^{2}_{W}}\right]\frac{C^{\varphi
  WB}}{\Lambda^{2}} %\nonumber \\&\quad
+ \left [ 0.16 - 0.22 \log
    \frac{\mu^{2}}{M^{2}_{W}}\right]\frac{C^{W}}{\Lambda^{2}}
    \nonumber \\[2mm]&\quad
+ \left [ 2.11 - 0.84 \log
    \frac{\mu^{2}}{M_W^2}\right]\frac{C^{uB}_{33}}{\Lambda^{2}}
  + \left [ 1.13 - 0.45 \log
    \frac{\mu^{2}}{M_W^2}\right]\frac{C^{uW}_{33}}{\Lambda^{2}}
    \nonumber \\[2mm]&\quad
- \left [ 0.03 + 0.01 \log
    \frac{\mu^{2}}{M_W^2}\right]\frac{C^{uB}_{22}}{\Lambda^{2}}
  - \left [ 0.01 + 0.00 \log
    \frac{\mu^{2}}{M_W^2}\right]\frac{C^{uW}_{22}}{\Lambda^{2}}
    \nonumber \\[2mm]&\quad
+ \left [ 0.03 + 0.01 \log
    \frac{\mu^{2}}{M_W^2}\right]\frac{C^{dB}_{33}}{\Lambda^{2}}
  - \left [ 0.02 + 0.01 \log
    \frac{\mu^{2}}{M_W^2}\right]\frac{C^{dW}_{33}}{\Lambda^{2}}
    \nonumber \\[2mm]&\quad
+ \left [ 0.02 + 0.00 \log
    \frac{\mu^{2}}{M_W^2}\right]\frac{C^{eB}_{33}}{\Lambda^{2}}
  - \left [ 0.01 + 0.00 \log
    \frac{\mu^{2}}{M_W^2}\right]\frac{C^{eW}_{33}}{\Lambda^{2}}
  + \ldots \,.
\end{align}
%%%%%%%%%%%%%%%
One can draw interesting remarks by comparing eqs.~\eqref{eq:num2} and
\eqref{eq:h2gg-num}. Wilson coefficients in the first line of both equations
are dominated from input scheme dependencies.\footnote{The large scheme
dependence can be understood by comparing \eqref{eq:h2gg-num} with the first
line of eq.~(5.1) of ref.~\cite{Dedes:2018seb}.} The only ``real'' difference is
a factor of 2 enhancement in front of the coefficient $C^{\varphi D}$ in the
case of $h\to \gamma\gamma$.
Another issue is the surprisingly large loop enhancement of the $C^{uB}_{33}$
coefficient (top-quark inside the loop) in $\delta \mathcal{R}_{h\to
\gamma\gamma}$ as shown and discussed in ref.~\cite{Dedes:2018seb}. This
enhancement has been reduced by a factor of 20 in $\delta \mathcal{R}_{h\to
Z\gamma}$ in \eqref{eq:num2}. The reason seems to be an accidental cancellation. 
In the $h\to \gamma\gamma$ case we have an overall factor $16 s c^{2}
\approx 6 $, while in the $h\to Z\gamma$ case we have an overall
factor $3c^{3} - 13 c s^{2}\approx -0.5$. It is this factor of $(-10)$
that gives such a big difference in the relevant results. This suppression may
be used to disentangle new-physics effects between the two observables.
Interestingly, however, the coefficient $C^{uW}_{33}$ does not suffer by similar
accidental suppression.
 
In comparing eqs.~\eqref{eq:num2} and \eqref{eq:h2gg-num}, even the dominant
contributions from the operators $C^{\varphi B}$ and $C^{\varphi WB}$ are
smaller in $h\to Z\gamma$ by factor of 3 and only the coefficient of $C^{\varphi
W}$ is similar in both $\delta \mathcal{R}_{h\to Z\gamma}$ and $\delta
\mathcal{R}_{h\to \gamma\gamma}$. This is very interesting for disentangling
among the three operators in case  new physics enters through those. For
example, one may envisage a new-physics scenario, like the one of
ref.~\cite{Bilenky:1993bt}, with a heavy hypercharged $SU(2)_L$-singlet
scalar, which is decoupled from the theory at the TeV scale. Since this will
only make $C^{\varphi B}$ non-zero, and say positive in $h\to \gamma\gamma$, it
will only make a suppressed reduction in case of $h \to Z \gamma$. However,
there are Wilson coefficients like the prefactor of $C^W$ that are similar in
both cases. 
 
The real power, however, of EFT, is when using experimental data to constrain
Wilson coefficients of various operators and therefore making estimates for
projective reach of observables. For example, bounds have been set in some of
the coefficients appearing in $\delta \mathcal{R}_{h\to \gamma\gamma}$ in
refs.~\cite{Dedes:2018seb, Dawson:2018liq}. We easily see that, if we consider
\emph{one coupling at a time}, bounds from $h\to \gamma\gamma$ on these
coefficients  kill any possible excess arising  from these operators in the
$h\to Z\gamma$ process. In addition to the already mentioned cancellation in
top-quark loop for $h  \to Z \gamma$, the relevant operators bounded from $h\to
\gamma\gamma$ are now numerically completely irrelevant for $h\to Z\gamma$.

That is quite a lot one can infer by just comparing only two observables! One
may use best fit values to EW observables to check upon other coefficients, such
as $C^W$, $C^{ll}_{1221}$, $C^{\varphi l (3)}$, $C^{\varphi u}$, $C^{\varphi
q(1,3)}$, $C^{\varphi D}$ that enter similarly in $\delta \mathcal{R}_{h\to
Z\gamma}$ and $\delta \mathcal{R}_{h\to \gamma\gamma}$ of eqs.~\eqref{eq:num2}
and \eqref{eq:h2gg-num}, respectively. By taking, for instance, the best fit
values from the 4th column of Table~6 in
ref.~\cite{Brivio:2017bnu}\footnote{Similar results one can draw from other
tree-level studies, see refs.~\cite{Ellis:2018gqa,Murphy:2017omb}.} we obtain
that it is unlikely to discover any possible new-physics effect through $h\to
Z\gamma$ decay in current LHC data before seeing a $h\to \gamma\gamma$ anomaly.
Of course this statement weakens if one allows for more operators to be present
at the same time. 

As we already mentioned in Introduction, below eq.~\eqref{mu}, in deriving bounds
from $\delta \mathcal{R}_{h\to Z \gamma}$ we implicitly assumed that at least
the dominant Higgs-boson production mechanism (gluon fusion) is not affected by
the operators involved in the $h\to Z\gamma$ decay. Indeed, for the same reason
we explained in section~\ref{sec:operators}, only CP-invariant operators
contribute to $gg\to h$ process. The main, i.e.\ tree level in SMEFT, gluon-higgs
operator, $Q_{\varphi G}$, as well as the ones affecting the one-loop diagrams,
$Q_{uG}$, $Q_{dG}$, do not interfere with the list of operators in
Table~\ref{tab:ops} relevant to $h\to Z\gamma$. However, operators
$Q_{u\varphi}$ and $Q_{d\varphi}$  enter in both $h\to Z\gamma$ and $h\to gg$ but
their associated Wilson coefficients are multiplied by small numbers in $\delta
\mathcal{R}_{h\to Z \gamma}$ of eq.~\eqref{eq:num2}. Finally, the combination
$(C^{\varphi\Box}+1/4 \, C^{\varphi D})$ enters only multiplicative in all
three observables, $h\to gg$, $h\to Z\gamma$ and $h\to \gamma\gamma$ which is
just a rescale effect. From these two coefficients, $C^{\varphi D}$ is a
custodial violating parameter and therefore highly suppressed. Of course the
safest is to calculate $h \to gg$ at one-loop in SMEFT. The reader is referred to
refs.~\cite{Maltoni:2016yxb,Cirigliano:2016nyn,Grazzini:2018eyk}

What about future $h\to Z\gamma$ sensitivity? Only at later stages of high
luminosity of $3000~\mathrm{fb}^{-1}$ at LHC, ATLAS will have enough
significance ($\sim 5\sigma$) for the  $h\to Z\gamma$
mode~\cite{Cepeda:2019klc}. Assuming SM Higgs production and decay, the signal
strength is expected to be measured with $\delta \mathcal{R}_{h\to Z \gamma}
\approx \pm 0.24$ uncertainty. On the other hand  for $h\to \gamma\gamma$, and
the same projective reach, ATLAS expects  $\delta \mathcal{R}_{h\to \gamma
\gamma} \approx \pm 0.04$ for a SM Higgs boson produced from gluon fusion
process and decaying dominantly  to $b\bar{b}$. By comparing, our EFT
calculations for $\delta \mathcal{R}_{h\to Z \gamma}$ and $\delta
\mathcal{R}_{h\to \gamma \gamma}$ in eqs.~\eqref{eq:num2} and
\eqref{eq:h2gg-num}, we obtain that any new-physics signal for $h\to Z\gamma$
is unlikely to be seen at near future LHC upgrades without seeing new 
physics first at $h\to \gamma\gamma$ data.

\section{Epilogue}
\label{sec:epilogue}

We have performed a one-loop calculation for the Higgs-boson decay to a
$Z$-boson and a photon, $h\to Z \gamma$, in SMEFT with $d=6$ operators written
in Warsaw basis. We find a general formula for the amplitude \eqref{eq:master2}
which is finite, it respects the Ward-identities, and is gauge-fixing parameter
independent. We present our result in terms of the ratio $\delta
\mathcal{R}_{h\to Z \gamma}$ in eq.~\eqref{eq:num2} and compare this with the
previously calculated ratio $\delta \mathcal{R}_{h\to \gamma \gamma}$. We find
that, for most Wilson-coefficients, $\delta \mathcal{R}_{h\to Z \gamma}$ is less
sensitive to new physics than $\delta \mathcal{R}_{h\to \gamma \gamma}$. Some of
the operators entering in $h\to Z \gamma$, but \emph{not} in $h\to \gamma
\gamma$, can modify $\delta \mathcal{R}_{h\to Z \gamma}$ at a rate hardly
noticeable, currently or in the near future, at the LHC.

%%%%%%%%%%%%%%%%%5
\subsection*{Acknowledgements}
%%%%%%%%%%%%%%%%%%
We would like to thank Janusz Rosiek and Michael Paraskevas for useful
discussions and comments on our manuscript, as well as Sally Dawson and Pier
Paolo Giardino for useful discussions about comparisons with
ref.~\cite{Dawson:2018pyl}. KS would like to thank the Greek State Scholarships
Foundation (IKY) for full financial support through the Operational Programme
``Human Resources Development, Education and Lifelong Learning, 2014-2020''. LT
is supported by the Onassis Foundation --- Scholarship ID: G ZO 029-1/2018-2019.

\bibliography{h2Zg-SMEFT}{}
%\addcontentsline{toc}{section}{References}
\bibliographystyle{JHEP}

\end{fmffile}
\end{document}